# N-doped graphitic carbon materials hybridized with transition metals (compounds) for hydrogen evolution reaction: understanding the synergistic effect from atomistic level

Wei Pei, Si Zhou[*], Yizhen Bai, Jijun Zhao

*Key Laboratory of Materials Modification by Laser, Ion and Electron Beams (Dalian University of Technology), Ministry of Education, Dalian 116024, China*

**Abstract**

The hybrid nanostructures of nitrogen doped carbon materials and nonprecious transition metals are among the most promising electrocatalysts to replace noble metal catalysts for renewable energy applications. However, the fundamental principles governing the catalytic activity of such hybrid materials remain elusive. Herein, we systematically explore the electrocatalytic properties of transition metals, transition metal oxides and carbides substrates covered by nitrogen-doped graphitic sheets for hydrogen evolution reaction (HER). Our first-principles calculations show that the graphitic sheet is prominently activated by the nitrogen doping and the coordinate bond with metal (compound) substrate through intralayer and interlayer charge transfer. Such hybrid materials can provide optimal binding capability for HER catalysis with Tafel barrier down to 1.0 eV. The HER activity can be correlated to the C $p_z$ band center, which is in turn governed by the electronic coupling strength between the graphitic sheet and metal substrate, thus paving a way to rational design of graphitic carbon/transition metal hybrid electrocatalysts of high performance.

[*] Corresponding author. Tel: 86-0411-84706100. E-mail: sizhou@dlut.edu.cn (Si Zhou)

# 1. Introduction

Aggravated energy and environmental crisis has stimulated intensive research on exploiting clean and renewable energy sources [1]. Hydrogen, as one of the most promising alternatives to fossil fuels [2], can be produced by water electrolysis. Hydrogen evolution reaction (HER) is the cathodic half reaction of water splitting and plays a pivotal role on the efficiency of electrolyzers [3–7]. To date, platinum group metals are known as the most effective electrocatalysts for HER [8–10]. However, the high cost and scarcity prohibit their widespread utilization. Significant progresses have been made in developing alternative catalysts for HER with inexpensive and earth-abundant elements, such as carbon-based materials [11–14], transition metal compounds (e.g., chalcogenides [15–20], carbides [21, 22], phosphides [23–25]), composites of carbon and transition metal based materials [26–30] and metal-organic frameworks (MOF) [31, 32].

Owing to their abundant sources, tunable compositions and versatile structures, heteroatom-doped carbon materials have generated tremendous research interests as promising electrocatalysts. Essentially, the differences in electronegativity and atomic size between the heteroatoms and carbon can polarize the adjacent carbon atoms and give rise to catalytic activity. Based on first-principles calculations combined with proof-of-concept experiments, Qiao et al. established a HER activity trend for B, N, O, P and S doped graphene [14]. They showed that co-doping N and P atoms into graphene can largely enhance the electrocatalytic activity giving much lower HER overpotential and higher current density than those of the single-doped graphene samples. In parallel, Chen et al. synthesized N and S co-doped nanoporous graphene with geometrical defects [12], and achieved high HER activity competitive to that of $MoS_2$ nanosheets [33, 34], which is known as the best Pt-free HER catalysts [8]. However, for most of the nonmetallic atom doped carbon materials, their HER activity is still unsatisfactory with regard to Pt group metals [8, 35].

Hybridizing carbon materials and transition metals (compounds) can induce synergistic effects and remarkably boost the electrocatalytic performance of the composite materials. Particularly, covering carbon layers over metal surfaces does not only gain prominent catalytic activity via interfacial charge transfer, but also protect metals from dissolution in the acidic environment. To this end, a variety of core-shell structures of graphitic carbon materials and transition metals have been synthesized, such as Fe and Co nanoparticles embedded in N-doped carbon nanotubes [26, 36], CoNi and FeCo nanoalloy encapsulated by N-doped graphene shells [29, 30]. In these core-shell structures, the number of graphitic shells can be as thin as single layer to a few layers. These hybrid nanostructures exhibit long-term durability and high HER activity with overpotentials and Tafel slopes approaching those of the benchmark Pt catalyst. Up to now, however, the core metals are restricted to only a small number of elements (Fe, Co, Ni). The underlying mechanism governing the HER activity of such hybrid materials and the principles for selecting appropriate core metals remain elusive, which hinders the advance of synthesizing hybrid catalysts with optimal performance.

Here we exploit a series of transition metals (Co, Fe), transition metal oxides ($Co_3O_4$, $Fe_3O_4$) and transition metal carbides (TiC, WC, VC) covered by N-doped graphitic sheets for HER catalysis. In previous experiments, these transition metals (compounds) have been hybridized with graphitic carbon materials and exhibit enhanced electrocatalytic activity for HER, oxygen reduction reaction and oxygen evolution reaction [37–43]. By systematic first-principles calculations, we screen the candidate graphitic carbon/transition metal hybrid systems for HER catalysts using the free energy of formation for hydrogen adsorption and kinetic barrier for hydrogen desorption and evolution as criteria for HER activity. The electronic origin of the enhanced activity and the key parameters determining the HER activity of these hybrid materials are elucidated. These theoretical results help prescribe the principles for designing efficient carbon/metal hybrid electrocatalysts for renewable energy applications.

## 2. Computational method

Spin-polarized density functional theory (DFT) calculations were carried out by the Vienna ab initio simulation package (VASP) [44], using the planewave basis with an energy cutoff of 500 eV, the projector augmented wave pseudopotentials [45], and the generalized gradient approximation parameterized by Perdew, Burke, and Ernzerhof (GGA-PBE) for exchange-correlation functional [46]. Grimme's semiempirical DFT-D3 scheme of dispersion correction was adopted to describe the van der Waals (vdW) interactions in layered materials [47]. The Brillouin zones of the supercells were sampled by 4 × 4 × 1 uniform **k** point mesh. With fixed cell parameters, the model structures were fully optimized using the convergence criteria of $10^{-4}$ eV for the electronic energy and $10^{-2}$ eV/Å for the forces on each atom. The Hubbard-U correction was applied for better description of the localized *d*-electrons of Co and Fe in their oxides using effective U−J values of 3.30 and 3.61 eV [48, 49], respectively. The on-site charge transfer was evaluated by the Bader charge analysis [50].

We considered a series of transition metals (TM), transition metal oxides (TMO) and carbides (TMC), including Co(111), $Co_3O_4$(111), $Fe_3O_4$(111), TiC(111) and VC(111) of the *fcc* phase, Fe(110) of the *bcc* phase, and WC(0001) surfaces of the *hcp* phase. For simplicity, we adopted monolayer graphene supported by these metals (compounds) to model the graphitic carbon/transition metal hybrid materials. The transition metal (compound) substrates were mimicked by slab models with 3~8 atomic layers, vacuum space of 16 Å thickness in vertical direction and 10~15 Å for the lateral dimensions. The in-plane lattices of transition metals (compounds) were slightly compressed or stretched to fit that of graphene, giving lattice mismatch of 1~5% (see Table 1 and Table S2 for details). Then, one N atom was substituted into the graphene sheet, resulting in doping concentrations of 2.44~4.00 at.% that are close to the typical doping levels of graphitic carbon materials in experiments [14, 51]. It is generally accepted that HER in the acidic media proceeds through two primary steps: a proton-coupled electron transfer to the catalysts forming an adsorbed H* species (Volmer reaction), and subsequent desorption of H* to form a $H_2$ molecule via either

Heyrovsky or Tafel mechanism [8]. The catalytic activity for HER can be characterized by the free energy of formation for hydrogen adsorption:

$$\Delta G_{H*} = \Delta E_{H*} + \Delta ZPE - T\Delta S \qquad (1)$$

where $\Delta E_{H*}$, $\Delta ZPE$ and $\Delta S$ are the differences of DFT total energy, zero-point energy and entropy between the adsorbed H* and gaseous $H_2$ phases, respectively; $T$ is temperature. The values of $\Delta ZPE$ and $\Delta S$ were taken from the NIST-JANAF thermodynamics table for $H_2$ molecule and by calculating the vibrational frequencies for the H* species (Table S1) [52], respectively. An ideal HER catalyst has $\Delta G_{H*} = 0$, while positive (negative) $\Delta G_{H*}$ values deviated from zero indicate weak (strong) H* binding by the catalysts that are adverse to H* desorption (adsorption).

The kinetic barrier and transition state for H* desorption and $H_2$ formation were simulated by the climbing-image nudged elastic band (CI-NEB) method [53]. Five images were used to mimic the reaction path. The intermediate images were relaxed until the perpendicular forces were less than 0.02 eV/Å.

## 3. Results and discussion

We start from examining the HER activity of freestanding N-doped graphene (hereafter denoted as NG). The graphitic sheet binds weakly with H* species. The C atoms adjacent to the N dopant provide the strongest binding strength; however, $\Delta G_{H*}$ is still too high (0.67 eV) for HER catalysis (Fig. S1). To enhance the surface reactivity of the graphene basal plane, we exploit the Co(111) substrate, which has a commensurate in-plane lattice with that of graphene. Upon optimization, the C atoms are in on-top registry with the underlying Co substrate (Fig. 1a), in good accord with the experimental observation [54]. The N dopants can stay on either the top site or hollow site of the Co substrates, resulting in only minor differences in the structural and binding properties (see Table 1, 2 and Table S2, S3 for details). The interaction between the graphitic sheet and metal substrate can be characterized by the interfacial binding energy ($E_{bind}$) defined as:

$$E_{\text{bind}} = (E_{\text{total}} - E_{\text{NG}} - E_{\text{sub}})/N_c \tag{2}$$

where $E_{\text{total}}$, $E_{\text{NG}}$ and $E_{\text{sub}}$ are the energies of the NG/Co hybrid system, freestanding doped graphene and standalone Co(111) substrate, respectively; $N_c$ is the number of C atoms in the graphitic sheet. As shown by Table 1, the NG/Co hybrid system has interlayer distance of 2.17 Å and interfacial binding energy of −0.08 eV per C atom. Differential charge densities show prominent electron accumulation in the interfacial region, signifying the coordinate bond between the graphitic sheet and the underlying metal atoms (Fig. S3). The graphitic sheet remains almost flat with vertical buckling of 0.02 Å.

The binding capability of the NG/Co hybrids is remarkably enhanced with $\Delta G_{H^*}$ lower than that of freestanding N-doped graphene by up to 0.54 eV. The H* binding strength highly depends on the chemical environment of C sites. Generally speaking, the C atoms adjacent to the dopants and on the hollow site of Co substrate provide the strongest H* binding (site 1 in Fig. 1a), while the C atoms far away from the dopants and on the top site of Co substrate provide weaker H* binding (site 4 in Fig. 1a). Optimum binding strength for HER catalysis can be achieved in these hybrid systems with $\Delta G_{H^*} = 0.13$ eV, rather competitive to that of Pt ($\Delta G_{H^*} = -0.09$ eV from our calculations, see Fig. 2a). The active sites are the C atoms adjacent to the N dopant and on the hollow site of Co substrate. The C atoms far away from dopants have $\Delta G_{H^*}$ of 0.4~0.8 eV, which is much enhanced with regard to that of freestanding graphene (1.84 eV), but still insufficiently strong for HER catalysis. In comparison, the standalone Co(111) surface shows $\Delta G_{H^*} = -0.48$ eV from our calculations, giving too strong binding for HER catalysis. Thus, by covering the graphitic sheet on Co metal and taking advantage of their interfacial coupling, it is possible to modulate the surface binding strength and catalytic activity for HER.

To explore the effect of various metals (compounds) on the binding properties of the supported graphitic sheets, we consider N-doped graphene on Fe(110), $Co_3O_4$(111), $Fe_3O_4$(111), TiC(111), VC(111) and WC(0001) substrates, as displayed by Fig. 1 (hereafter denoted as NG/Fe, NG/$Co_3O_4$, NG/$Fe_3O_4$, NG/TiC, NG/VC and NG/WC, respectively). The interlayer distances of these hybrid systems range from 2.10 to 2.47 Å with interfacial binding energies of –0.36~–0.09 eV per C atom (Table 1). The graphitic sheets of NG/Fe$_3$O$_4$ and NG/WC hybrids exhibit large vertical buckling of 0.21~0.25 Å, while the carbon layer in the other hybrid systems remains almost flat (vertical buckling < 0.07 Å). Distinct from the Co(111) surface, the in-plane lattices of these substrates are not commensurate with that of graphene. Thus, for each hybrid system, we have considered two models with the graphitic sheet in different relative positions to the substrates. These models show very similar structural, electronic and binding properties, meaning that our theoretical results do not rely on the specific structural model (see Table 1, 2 and Table S2, S3 for details).

The binding properties of N-doped graphene on various substrates are presented in Table 2 and Fig. 2b. Overall speaking, the H* binding strength follows the sequence: NG/VC ~ NG/WC > NG/Fe > NG/Co > NG/Co$_3$O$_4$ > NG/Fe$_3$O$_4$ ~ NG/TiC. The strongest H* binding can be achieved on the C atoms adjacent to the N dopants and close to the hollow site of the underlying substrates. For the NG/VC and NG/WC systems, the lowest $\Delta G_{H^*}$ attains about –0.33 eV, too strong for H* desorption and H$_2$ formation; optimum binding for HER catalysis with $\Delta G_{H^*}$ = 0.06 eV can be achieved on the C atoms adjacent to N and close to the top site of substrates, as well as those in the meta-position of N dopants and close to the hollow site of substrates. For NG/VC, the hollow site C atoms far away from N give $\Delta G_{H^*}$ = 0.15~0.19 eV, which may also have certain HER activity. The NG/Fe system can provide moderate H* binding with $\Delta G_{H^*}$ = –0.02 eV, which is achieved on the hollow site C atoms adjacent to the N dopants. On the contrary, the NG/Co$_3$O$_4$, NG/Fe$_3$O$_4$ and NG/TiC systems exhibit weaker binding capability with the lowest $\Delta G_{H^*}$ of 0.24~0.37 eV, respectively, which are too weak to adsorb H* species and thus not eligible for HER catalysis.

The enhanced HER activity of the graphitic carbon/transition metal hybrid systems is attributed to the electron injection from the metal (compound) substrates to the graphitic sheet, which destructs the π conjugation and partially occupied the $p_z$ orbitals of C atoms. According to the Bader charge analysis, each C atom gains 0.02~0.05 electrons on average from the substrates. Consequently, the graphitic sheets exhibit metallic behavior with large amount of electronic states near the Fermi level, as illustrated by the density of states (DOS) (Fig. 3a and Fig. S4). The distinct binding capabilities of various hybrid systems can be correlated to the center of C $p_z$ bands ($\varepsilon_{pz}$) defined as

$$\varepsilon_{p_z} = \frac{\int_{-\infty}^{0} E D(E) \, dE}{\int_{-\infty}^{0} D(E) \, dE} \tag{3}$$

where $D(E)$ is the DOS of C $p_z$ band at a given energy $E$. Fig. 3b shows a clear linear relationship between $\varepsilon_{pz}$ and the lowest $\Delta G_{H^*}$ for various NG/metal systems. The hybrid system with lower $p_z$ band center yields stronger H* binding, in contrary to the well-known trend of $d$-band center vs. H* binding for transition metals (compounds) [55]. This distinct behavior is ascribed to the fully occupancy of antibonding states ($\sigma^*$) between H* and graphitic sheet, as demonstrated by Fig. 3c and Fig. S5 that few electron states are available in the conduction bands. As a result, deeper valence orbital levels of the graphitic sheet lead to stronger bonding with H* species according to the extended Hückel theory (see Supplementary Data for details) [56]. Note that the NG/WC system is a little deviated from the linear relation between $\varepsilon_{pz}$ and $\Delta G_{H^*}$ in Fig. 3b. This may be due to the fact that $\varepsilon_{pz}$ is correlated to the overall surface binding capability of the considered hybrid systems. The graphitic sheet in NG/WC experiences severer vertical buckling (0.25 Å) than the other hybrids, which may prominently impact the $\Delta G_{H^*}$ values of specific C sites.

Moreover, the N dopants in the graphitic sheet induce electron redistributions, which together with the interfacial charge transfer dictate the binding strength of various C sites within a hybrid system. The C atoms carrying less electrons provide stronger H* binding, as illustrated by Fig. 4. In particular, the C atoms on the top site of metal substrate gain more electrons from the underlying metal atoms than those on the hollow site, the C atoms adjacent to the N dopants donate electrons to N, and consequently, the binding capability acends for these C sites (Table 2 and Table S3).

Therefore, by choosing appropriate metal (compound) substrates coorperated with N-doping, it is possible to modulate the electronic states and electron density distributions of the graphitic carbon materials, and ultimately achieving the optimal surface binding properties for HER catalysis.

To further evaluate the catalytic performance of these NG/metal hybrid systems, we calculated the kinetic barriers for two adsorbed H* species on the active sites of NG/Co, NG/Fe, NG/WC and NG/VC to form a $H_2$ molecule under the Tafel mechanism. The transition states involve two adjacent H atoms staying above the carbon surfaces at about 1.1 Å vertical distance (Fig. 5). The obtained energy barriers are 1.47, 1.93, 1.01, and 1.42 eV for the NG/Co, NG/Fe, NG/WC and NG/VC systems, respectively, much lowered than the values of freestanding carbon-based materials (~2.0 eV) [14], and comparable to those of the modified $MoS_2$ (1.0~1.5 eV) [18]. Therefore, the proposed graphene/metal hybrid nanostructures can provide suitable binding strength for HER catalysis from a thermodynamic point of view, and their catalytic activities are mainly limited by the kinetic process of H* desorption and $H_2$ formation.

## 4. Conclusions

In summary, we systematically investigate the electrocatalytic properties of N-doped graphitic carbon materials hybridized with transition metals (Co, Fe), transition metal oxides ($Co_3O_4$, $Fe_3O_4$) and carbides (TiC, WC, VC) for hydrogen evolution reaction. Our first-principles calculations show that the synergy of N doping and substrate interaction significantly boosts the reactivity of graphene basal plane by intralayer and interlayer charge transfer. The doped graphitic sheet supported by Co, Fe, WC and VC substrates can provide moderate binding with H* species suitable for HER catalysis, with $\Delta G_{H*}$ close to zero and Tafel barrier down to 1.0 eV. The active sites are the C atoms close to N dopants. The C $p_z$ band center, governed by the electron coupling strength between the graphitic sheet and metal (compound) substrates, can serve as a descriptor for modulating the H* binding strength and thus

HER activity for the graphitic carbon/transition metal hybrids. These theoretical results provide vital guidance for compositing carbon and transition metal based materials with optimal catalytic performance for HER as well as other key electrochemical reactions in energy conversion technology.

**Acknowledgments**

This work was supported by the National Natural Science Foundation of China (11504041, 11574040), the Fundamental Research Funds for the Central Universities of China (DUT16LAB01, DUT17LAB19), and the Supercomputing Center of Dalian University of Technology.

**Appendix A. Supplementary data**

Supplementary data related to this article is available.

**Table 1.** Structural and electronic properties of N-doped graphene on various transition metal-based substrates (as shown in Fig. 1), including lattice mismatch ($\delta$), interlayer distance ($d$), vertical buckling of the graphitic sheet ($\Delta d$), interfacial binding energy per C atom ($E_{bind}$), and C $p_z$ band center ($\varepsilon_{pz}$).

| structure | $\delta$ | $d$ (Å) | $\Delta d$ (Å) | $E_{bind}$ (eV) | $\varepsilon_{pz}$ (eV) |
|---|---|---|---|---|---|
| NG/Co | 1.99% | 2.17 | 0.04 | −0.08 | −5.12 |
| NG/Fe | 3.81% | 2.10 | 0.05 | −0.09 | −5.28 |
| NG/Co$_3$O$_4$ | 0.84% | 2.18 | 0.07 | −0.13 | −4.99 |
| NG/Fe$_3$O$_4$ | 4.75% | 2.14 | 0.21 | −0.33 | −4.77 |
| NG/TiC | 3.18% | 2.24 | 0.07 | −0.37 | −4.96 |
| NG/WC | 2.70% | 2.47 | 0.25 | −0.13 | −5.26 |
| NG/VC | 4.22% | 2.12 | 0.06 | −0.32 | −5.43 |

**Table 2.** Free energy of formation for hydrogen adsorption ($\Delta G_{H^*}$) and number of electrons (CT) by Bader charge analysis for various C sites of N-doped graphene on transition metal (compound) substrates (as shown in Fig. 1).

| system | site 1 | | site 2 | | site 3 | | site 4 | |
|---|---|---|---|---|---|---|---|---|
| | $\Delta G_{H^*}$ (eV) | CT ($e$) | $\Delta G_{H^*}$ (eV) | CT ($e$) | $\Delta G_{H^*}$ (eV) | CT ($e$) | $\Delta G_{H^*}$ (eV) | CT ($e$) |
| NG/Co | 0.13 | 0.13 | 1.01 | –0.32 | 0.82 | –0.30 | 0.42 | –0.27 |
| NG/Fe | –0.02 | 0.25 | 0.61 | –0.21 | 0.40 | –0.20 | 0.32 | –0.14 |
| NG/Co$_3$O$_4$ | 0.24 | 0.18 | 0.52 | –0.04 | 0.58 | –0.11 | 0.32 | 0 |
| NG/Fe$_3$O$_4$ | 0.37 | –0.05 | 0.60 | –0.07 | 0.93 | –0.12 | 0.89 | –0.10 |
| NG/TiC | 0.34 | –0.11 | 0.71 | –0.14 | 0.87 | –0.23 | 0.84 | –0.22 |
| NG/WC | 0.07 | 0.31 | 0.18 | –0.03 | 0.36 | –0.11 | 0.25 | –0.08 |
| NG/VC | –0.26 | 0.22 | 0.06 | 0.04 | 0.50 | –0.17 | 0.19 | –0.04 |

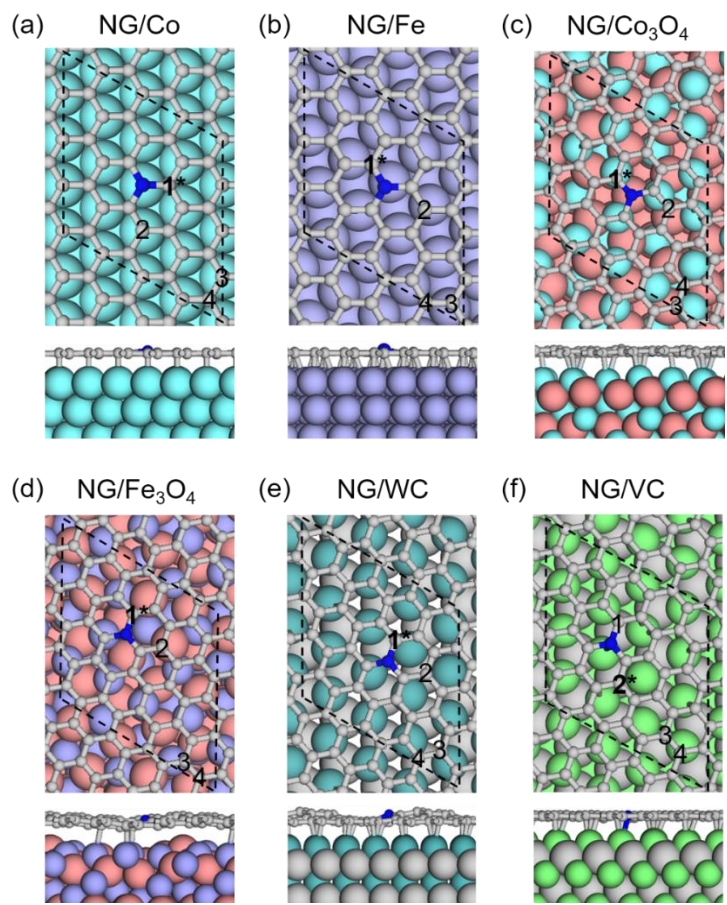

**Fig. 1**. Top and side views of N-doped graphene on (a) Co(111), (b) Fe(110), (c) $Co_3O_4$(111), (d) $Fe_3O_4$(111), (e) WC(0001), and (f) VC(111) substrates. The black boxes in the top views show the lateral dimensions of supercells. The numbers indicate the C sites whose $\Delta G_{H^*}$ values are calculated and reported in Table 2, and the star denotes the most active site for HER catalysis. The C, N, O, Co, Fe, W and V atoms are shown in grey, blue, coral, cyan, purple, turquoise and green colors, respectively.

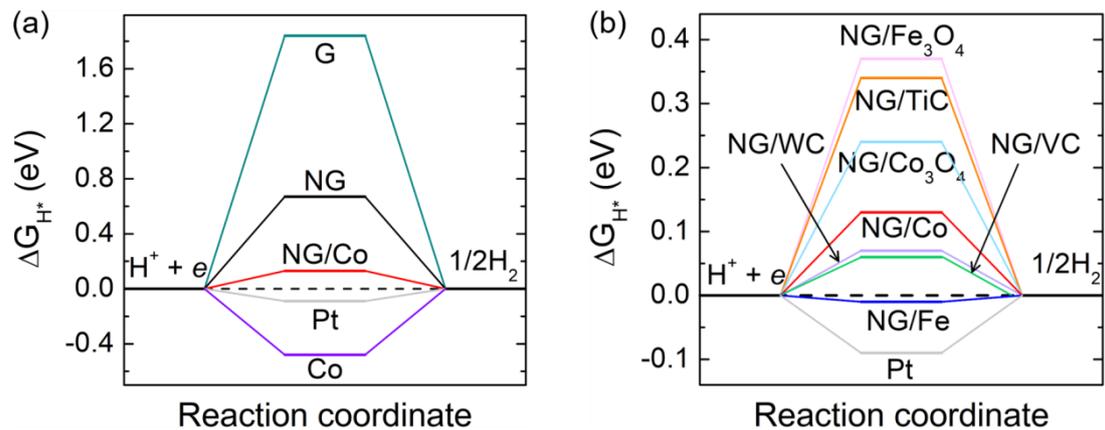

**Fig. 2.** Free energy diagrams for hydrogen evolution at zero potential and pH = 0 on (a) freestanding graphene (G) and N-doped graphene (NG), standalone Co(111) and Pt(111) surfaces, and the NG/Co hybrid system, and (b) N-doped graphene on various transition metal (compound) substrates. The black dashed line indicates the ideal $\Delta G_{H^*}$ for HER catalysis.

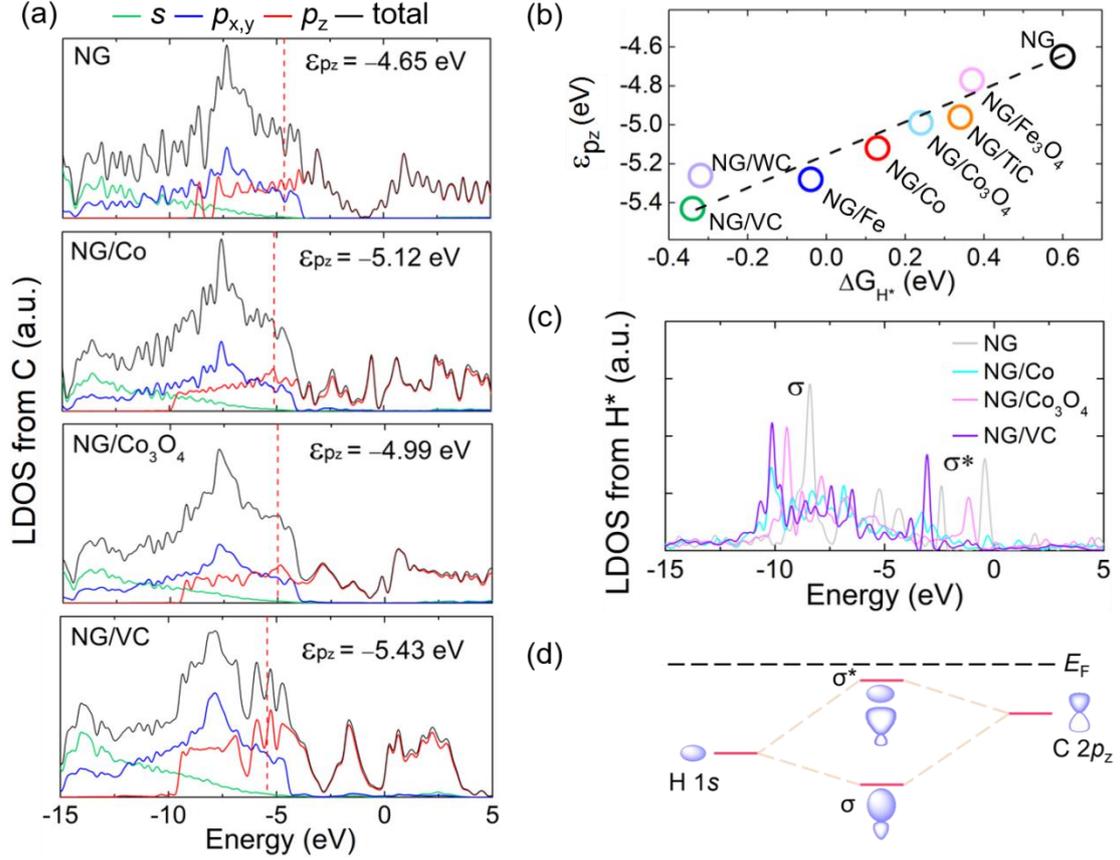

**Fig. 3.** (a) Local density of states (LDOS) of C atoms in freestanding N-doped graphene (NG), and the NG/Co, NG/Co$_3$O$_4$ and NG/VC hybrid systems. The colored solid lines show the projected DOS from $s$, $p_{x,y}$ and $p_z$ orbitals, respectively. The red dashed lines and the numbers next to them indicate the C $p_z$ band centers for each system. (b) The C $p_z$ band center as a function of the lowest $\Delta G_{H^*}$ value for NG and various NG/metal hybrid systems. (c) LDOS of a H atom adsorbed on the most active site of NG and the NG/Co, NG/Co$_3$O$_4$ and NG/VC hybrid systems. (d) Schematic illustration of formation of H−C bonds for the NG/metal hybrid systems. The bonding and antibonding states are indicated by σ and σ*, respectively.

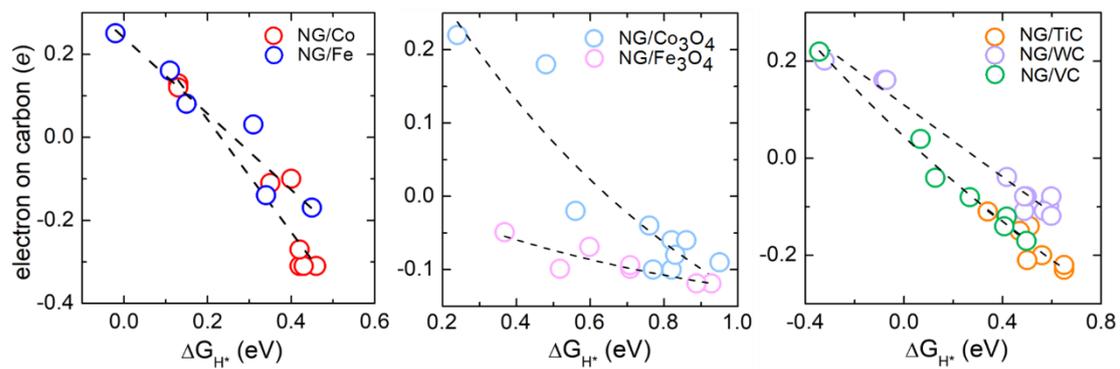

**Fig. 4.** Number of electrons on C atoms vs. lowest $\Delta G_{H*}$ values for N-doped graphene on various transition metal (compound) substrates. The black dashed lines are guide for eyes.

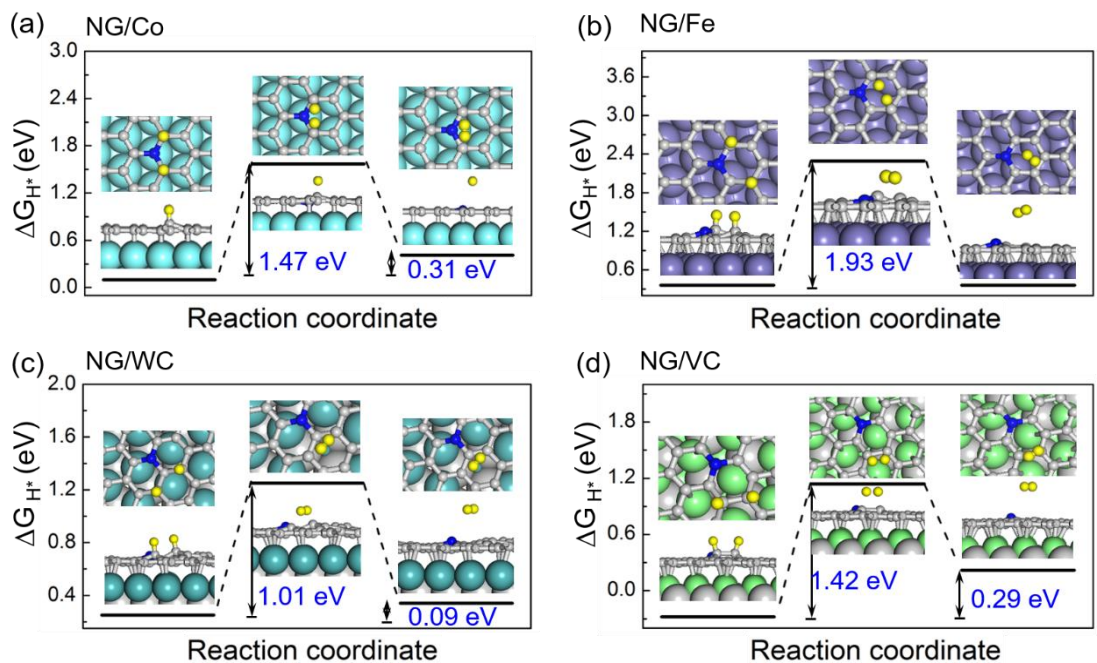

**Fig. 5.** Free energy profiles of Tafel mechanism pathway for hydrogen evolution on (a) NG/Co, (b) NG/Fe, (c) NG/WC and (d) NG/VC hybrid systems. The blue numbers (from left to right) indicate the kinetic barriers and Gibbs free energy of formation, respectively. The H, C, N, V, Fe, Co and W atoms are shown in yellow, grey, blue, green, purple, cyan and turquoise colors, respectively.